\documentclass[12pt,fullpage]{article}

\def\be{\begin{equation}}
\def\ee{\end{equation}}
\def\bea{\begin{eqnarray}}
\def\eea{\end{eqnarray}}

\begin{document}

\begin{flushright}
TIFR/TH/04-02
\end{flushright}
\bigskip
\begin{center}
{\large N-PARTITE SEPARABILITY INEQUALITIES
EXPONENTIALLY STRONGER THAN LOCAL REALITY INEQUALITIES} \\[8mm]
S.M. Roy$^\star$ \\
Department of Theoretical Physics \\
Tata Institute of Fundamental Research \\
Homi Bhabha Road, Colaba, Mumbai - 400 005, India 
\end{center}

\vfill

\noindent Abstract.  For an $N$-partite quantum
system we show that 
separability implies inequalities on Bell correlations which are
stronger than the local reality inequalities by a factor
$2^{(N-1)/2}$.

\vfill


\vfill

\noindent $^\star$ E-mail: shasanka@theory.tifr.res.in

\newpage

\noindent 1. \underbar{Introduction}: Striking features of quantum
entanglement were brought into sharp focus by the landmark papers of
Einstein, Podolsky and Rosen and Schr\"odinger$^1$.  The Bohm-Aharonov
version of the EPR paradox with two spin half particles in a singlet
state led to Bell's theorem$^2$ that quantum theory violates `local
realism'.  Subsequently other entangled states,
e.g. Greenberger-Horne-Zeilinger$^3$ (GHZ) state of four spin
${1\over2}$ particles and its $N$-particle generalizations led to
optimal local reality inequalities derived by Mermin$^4$ (for odd
$N$) and Roy and Singh$^5$ (all $N$) and others$^6$.  The
Mermin-Roy-Singh (MRS) inequalities are violated by suitable entangled
or non-separable states by a factor $2^{(N-1)/2}$.  This is now
recognised to be the maximal violation possible, due to Cirel'son's
theorem$^7$ for $N=2$, and recent work of Werner and Wolf$^8$ for
general $N$.

The physical origins of the ideas of local realism and quantum
separability are very different.  The exponential violation of local
realism by suitable entangled states is therefore an extremely
interesting but indirect consequence of entanglement.  The current
explosive interest in applications of quantum entanglement to quantum
information theory$^9$ prompts us to seek direct qualitative and
quantitative signatures of quantum entanglement.  Here we present
separability inequalities on Bell correlations which are
exponentially stronger than local reality inequalities for large
$N$.  We show that for
 $N$-partite systems there are entangled states
violating separability by a factor  $2^{(N-1)}$.  It
is natural to conjecture that this exponential violation of
separability, being a quantitative measure of quantum parallelism, is
intimately connected to the exponential speed-ups achievable in
quantum computation$^{10}$.
\bigskip

\noindent 2. \underbar{Local Realism Versus Separability}:  Consider a
composite system which breaks up into  $N$
components.  The  $k$th component is measured with
apparatus specified by a set of parameters  $a_k$ to
determine the value of a variable  $A^{(k)} (a_k)$
which by its very definition must lie between
 $\nu_k$ and  $\mu_k$, 
\[
\nu_k \leq A^{(k)} (a_k) \leq \mu_k.
\]
Repeated simultaneous measurements of  $A^{(k)}
(a_k)$ yield their correlation function as the expectation
value  $\langle A(a)\rangle$, where 
\be
A(a) = \prod^N_{k=1} A^{(k)} (a_k),
\ee
and  $a \equiv (a_1,a_2,\cdots,a_N)$.  A Bell
correlation function  $\langle B\rangle$ is a linear
combination of such correlation functions, with the Bell variable
$B$ defined by
\be
B = \sum_a c(a) A(a),
\ee
where  $c(a)$ are real numbers.

According to Bell's formulation of Einstein locality or local realism,
if all pairs made out of the  $N$ sub-systems are
mutually spacelike separated, then in a `Local Hidden Variable'
theory,
\be
\langle A(a) \rangle_{LHV} = \int d\lambda \rho(\lambda) \prod^N_{k=1}
A^{(k)} (\lambda,a_k),
\ee
where `$\lambda$' are hidden variables which
determine the outcomes  $A^{(k)} (\lambda, a_k)$ in
individual runs, and  $\rho (\lambda)$ their
probability distribution.  Local reality means that for each
 $\lambda$ the outcome  $A^{(k)}
(\lambda,a_k)$ is independent of all other orientations
 $a_\ell$ and outcomes  $A^{(\ell)}
(\lambda,a_\ell)$ for  $\ell \neq k$ observed at
spacelike separation.

On the other hand in quantum theory each  $A^{(k)}
(a_k)$ becomes a self-adjoint operator in a Hilbert space
${\cal H}^{(k)}$ with eigenvalues in the interval
$[\nu_k,\mu_k]$, and
\be
\langle A(a) \rangle = {\rm Tr} \ \rho \prod^N_{k=1} A^{(k)} (a_k), 
\ee
where $\rho$ is the density operator for the quantum
state defined on  $\displaystyle{\bigotimes_{k=1}^N} {\cal H}^k$.

\underbar{In defining quantum separability, there is no reference to 
spatial} \\ \underbar{separation of the subsystems}.  For pure states
of bipartite 
systems, $\rho = |\psi\rangle \langle \psi|$, where any
$\psi$ can be written as a Schmidt biorthogonal
sum$^{11}$
\be
\psi = \sum^M_{i=1} \sqrt{p_i} \psi_i^{(1)} \psi_i^{(2)},  
\ee
with $p_i > 0$,
$\displaystyle{\sum_i} p_i = 1$.  The state is
called separable if the Schmidt rank  $M=1$ and
called entangled or EPR correlated if  $M > 1$.  For
 $N$-partite systems with  $N >
2$, and for mixed states of bipartite systems we must use the
following more general definition$^{12,13}$.  A density operator
 $\rho$ on the tensor product of
 $N$ Hilbert spaces
 $\displaystyle{\bigotimes^N_{k=1}} {\cal H}^{(k)}$ is called
separable or disentangled or classically correlated if it can be
written as a convex combination of tensor product states
\be
\rho = \sum_i r_i \bigotimes^N_{k=1} \rho_i^{(k)}
\ee
(where the sum converges in trace class norm), with
$r_i > 0$,  ${\displaystyle{\sum_i}}
r_i = 1$.  Otherwise it is called entangled or EPR correlated.

Whereas local realism is a concept independent of quantum theory,
separability is formulated entirely in terms of quantum theory.  What
is the precise connection between them?  Consider first
 $N=2$.  The LHV representationimplies the Bell-CHSH
inequalities, but it is known$^{14}$ that these inequalities are not
sufficient to derive the LHV representation.  Fortunately, it is easy
to show$^{12}$, without going via Bell inequalities, that for
separable quantum states, the Bell correlations
 $A(a)$ obey a LHV representation.  For separable
states 
\be
{\rm Tr} \rho A^{(1)} (a_1) A^{(2)} (a_2) \cdots A^{(N)} (a_n)
= \sum_i r_i A^{(1)} (a_1,i) A^{(2)} (a_2,i)
\cdots A^{(N)} (a_n,i), 
\ee
where  $A^{(k)} (a_k)$ are observables on
 ${\cal H}^{(k)}$ depending on parameter sets
 $a_k$, and
\be
A^{(k)} (a_k,i) = {\rm Tr} \rho_i^{(k)} A^{(k)} (a_k)
. 
\ee
The decomposition (7) is exactly of the Bell Local Hidden Variables
(LHV) form and readily shows that all Bell correlations in separable
states must obey the Bell local-realism inequalities.  There exist
partial results in the reverse direction.  Gisin and Peres$^{15}$
showed using the Schmidt decomposition that for every pure entangled
state of a bipartite system one can find observables whose Bell
correlations violate local realism inequalities.  However,
Werner$^{12}$ constructed a class of mixed entangled states for
 $N=2$ which nevertheless admit a LHV representation
for Bell correlations of all observables.  Thus there is no one to one
correspondence between separable quantum states and those admitting a
LHV representation for Bell correlations.  The focus of the present
work will be to show that maximal violations of separability
inequalities for  $N$-partite systems are
exponentially higher than the maximal violations of local reality.
\bigskip

\noindent 3. \underbar{Summary of MRS Local Reality Inequalities For 
$N$ qubit systems}:  Consider the 
operators  $A^{(k)} (a_k)$ to be
 $\sigma_x^{(k)}$ or
 $\sigma_y^{(k)}$, the Pauli spin operators for the
 $k$th qubit.  Define
 $\sigma_\pm^{(k)} = \sigma_x^{(k)} \pm
i\sigma_y^{(k)}$, and the Bell operators 
\bea
B_+ &=& {1\over2} \left(\bigotimes^N_{k=1} \sigma_+^{(k)} + \bigotimes^N_{k=1}
\sigma_-^{(k)} \right) \\
B_- &=& {1\over 2i} \left(\bigotimes^N_{k=1} \sigma_+^{(k)} -
\bigotimes^N_{k=1} \sigma_-^{(k)} \right)
\eea
which are of the general form given by (1) and (2) when reexpressed in terms of
 $\sigma_x^{(k)},\sigma_y^{(k)}$.
Can the quantum Bell correlations which are expectation values of the
operators  $B_\pm$ be reproduced by the
corresponding LHV expressions,
\be
\langle B_+\rangle_{\rm LHV} = \Re e \int d\lambda \rho(\lambda) 
\prod^N_{k=1} \left(\sigma_x^{(k)} (\lambda) + i\sigma_y^{(k)}
(\lambda) \right) 
\ee
\be 
\langle B_- \rangle_{\rm LHV} = \Im m \int d\lambda \rho(\lambda) 
\prod^N_{k=1} \left(\sigma_x^{(k)} (\lambda) + i\sigma_y^{(k)}
(\lambda)\right), 
\ee
where  $-1 \leq \sigma_{x,y}^{(k)} (\lambda) \leq
1$?  The  $\langle B_\pm \rangle_{\rm LHV}$ are
linear in each  $\sigma_{x,y}^{(k)} (\lambda)$ and
their extreme values must be reached when
 $\sigma_{x,y}^{(k)} (\lambda) = \pm 1$.  The MRS
procedure quickly yields the bounds,
\be
\left| \langle B_\pm \rangle_{\rm LHV} \right| \leq 2^{(N-1)/2}, \ N \
{\rm odd} 
\ee
\be
\left| \langle B_+ \rangle_{\rm LHV} \right| + \left| \langle B_-
\rangle_{\rm LHV} \right| \leq 2^{N/2}, \ N \
{\rm even,}
\ee
which are known to be violated by quantum correlations by a factor
 $2^{(N-1)/2}$. 
\bigskip

\noindent 4. \underbar{Separability Inequalities For
 $N$ Qubit Systems}:  Consider first
the factorized state
\[
|\psi\rangle = \bigotimes^N_{k=1} |\psi^{(k)}\rangle, \ |\psi^{(k)}\rangle =
\left(\matrix{\alpha^{(k)} \cr \beta^{(k)}} \right)
\]
with $|\alpha^{(k)}|^2 + |\beta^{(k)}|^2 = 1$.  This yields
\be
\left|\langle \psi | \bigotimes^N_{k=1} \sigma_-^{(k)} |\psi\rangle \right| =
\left|\prod^N_{k=1} 2\beta^{(k)\star} \alpha^{(k)} \right| \leq 1
\ee
Hence, for the Bell correlations in factorized states, 
\[
|\langle \psi|B_\pm|\psi\rangle| \leq 1,
\]
\[
|\langle \psi | B_+|\psi\rangle | + |\langle \psi | B_-|\psi\rangle |
\leq \sqrt{2},
\]
where $B_\pm$ are the operators defined by Eqs. (9) and (10).  
We now show that the expection values of
 $B_\pm$ in 
a general separable state (6) must obey the same inequalities.  Each
density operator  $\rho_i^{(k)}$ in (6) is a convex combination of
pure states,  
\[
\rho_i^{(k)} = \sum_s c_{is}^{(k)} |\psi_{is}^{(k)}\rangle \langle
\psi_{is}^{(k)}|,
\]
with  $c_{is}^{(k)} > 0$, 
${\displaystyle 
\sum_{s}} c_{is}^{(k)} = 1$.  We readily deduce by using the convexity
properties and a relabelling of indices that a general separable
density operator (6) can also be written as a convex combination of
tensor products of pure states
\[
\rho = \sum_I r_I \bigotimes^N_{k=1} |\psi_I^{(k)} \rangle \langle
\psi_I^{(k)}|, 
\]
with $r_I > 0$, ${\displaystyle \sum_I} r_I = 1$.  Hence,
\bea
\left|{\rm Tr} \ \rho \bigotimes^N_{k=1} \sigma_-^{(k)}\right| &=&
\left|\sum_I r_I \prod^N_{k=1} \langle \psi_I^{(k)} |\sigma_-^{(k)}|
\psi_I^{(k)}\rangle\right| \nonumber \\
\leq \sum_I r_I &=& 1
\eea
where we have used the positivity of $r_I$ and the result (15) for
factorized states.  This immediately implies that the Bell
correlations in arbitrary separable states must obey
\be
|{\rm Tr} \ \rho B_\pm| \leq 1,
\ee
and
\be
|{\rm Tr} \ \rho B_+| + |{\rm Tr} \ \rho B_-| \leq \sqrt{2}, 
\ee
for every  $N$-partite separable density operator
$\rho$.  These are the announced ``Separability
inequalities''.  Comparison with Eqs. (13), (14) shows that the
separability inequalities are stronger than the local reality
inequalities by a factor  $2^{(N-1)/2}$.  We expect the experimental
violations of (17), (18) by entangled states to be useful signatures
of non-separability.  We also hope that a theoretical study of these
exponential violations will illuminate the mechanism of speed-up
achieved in quantum computation.
\bigskip

\noindent 5. \underbar{Quantum Violations of Separability and Local
Reality}: For the entangled pure state 
\[
|\psi\rangle = {1 \over \sqrt{2}} (\uparrow\uparrow \cdots \uparrow +
e^{i\theta} \downarrow\downarrow \cdots \downarrow),
\]
where  $\uparrow$ and
 $\downarrow$ denote eigenstates of
 $\sigma_z$ with eigenvalues  $\pm 1$, we have,
\be
\langle \psi | B_+ |\psi\rangle = \cos\theta \ 2^{N-1}
\ee
\be
\langle \psi |B_- |\psi\rangle = \sin\theta \ 2^{N-1}. 
\ee
Taking successively  $\theta = 0,\pi/2$
and  $\pi/4$ we see that the values of
 $\langle B_+\rangle$,  $\langle
B_-\rangle$ and  $|\langle B_+\rangle | + |\langle
B_- \rangle |$ violate the separability inequalities by a factor
$2^{N-1}$, and the local reality inequalities by a
factor  $2^{(N-1)/2}$.  We can prove that these
violations are maximal by a generalization of Cirel'son's theorem to
$N$-partite systems$^{16}$ or by the variational
methods of Werner and Wolf$^8$.

The present work can be generalized in several directions.  The
inequalities can be generalized to the case of partially separable
density matrices of  $N$-partite systems and
compared with partial local hidden variable theory inequalities
obtained by Svetlichny and others$^{17}$.  We have also obtained
separability inequalities for  $N$-Qudit
systems$^{18}$. 

I am grateful to Virendra Singh for many collaborations on Bell
inequalities and a careful reading of this manuscript.
\bigskip\bigskip

\noindent References
\begin{enumerate}
\item[{1.}] A. Einstein, B. Podolsky and N. Rosen,
Phys. Rev. \underbar{47}, 777 (1935), E. Schr\"odinger, Naturwiss
\underbar{23}, 807,823,844 (1935).
\item[{2.}] J.S. Bell, Physics (Long Island City, N.Y.) \underbar{1},
195 (1964); J.F. Clauser, M.A. Horne, A. Shimony, and R.A. Holt,
Phys. Rev. Lett. \underbar{26}, 880 (1969); J.S. Bell, in
``Foundations of Quantum Mechanics'', International School of Physics
``Enrico Fermi'', Course IL (Academic, New York, 1971), p.171.
\item[{3.}] D.M. Greenberger, M.A. Horne and A. Zeilinger, in ``Bell's
Theorem, Quantum Theory and Conceptions of the Universe'', edited by
M. Kafatos (Kluwer Academic, Dordrecht, 1989), p.69.
\item[{4.}] N.D. Mermin, Phys. Rev. Lett. \underbar{65}, 1838 (1990).
\item[{5.}] S.M. Roy and V. Singh, Phys. Rev. Lett. \underbar{67},
2761 (1991).
\item[{6.}] M. Ardehalli, Phys. Rev. A\underbar{46}, 5375 (1992);
A.V. Belinskii and D.N. Klyshko, Phys. Usp. \underbar{36}, 653 (1993);
N. Gisin and H. Bechmann-Pasquinucci, Phys. Lett. A\underbar{246}, 1
(1998). 
\item[{7.}] B.S. Cirel'son, Lett. Math. Phys. \underbar{4}, 93 (1980);
L.J. Landau, Phys. Lett. A\underbar{120}, 54 (1987).
\item[{8.}] R.F. Werner and M.M. Wolf, Phys. Rev. A\underbar{64},
032112 (2001).
\item[{9.}] ``Special Issue: Quantum Information Theory'',
Journ. Math. Phys. \underbar{43}, No. 9, 4237-4570 (2002). 
\item[{10.}] P. Shor, Proc. of 35th Annual Symposium on the
Foundations of Computer Science, (JEEE Computer Society, Los
Alamitos), p.124 (1994); quant-ph/9508027, and SIAM Journal on
Computing \underbar{26}, 1484 (1997).
\item[{11.}] E. Schmidt, Math. Annalen \underbar{63}, 433 (1906);
A. Ekert and P.L. Knight, Am. J. Phys. \underbar{63}, 415 (1995).
\item[{12.}] R.F. Werner, Phys. Rev. A\underbar{40}, 4277 (1989).
\item[{13.}] M.J. Donald, M. Horodecki and O. Rudolph, in Ref. 9,
p.4252. 
\item[{14.}] S.M. Roy and V. Singh, J. Phys. A,
Math. Gen. \underbar{11}, L167 (1978); A. Garg and N.D. Mermin,
Phys. Rev. Lett. \underbar{49}, 1220 (1982).
\item[{15.}] N. Gisin and A. Peres, Phys. Lett. A\underbar{162}, 15
(1992).
\item[{16.}] S.M. Roy, `Cirel'son's theorem for Bell correlations of
N-partite systems', in preparation.
\item[{17.}] G. Svetlichny, Phys. Rev. D\underbar{35}, 3066 (1987);
M. Seevinck and G. Svetlichny, Phys. Rev. Lett. \underbar{89}, 060401
(2002). 
\item[{18.}] S.M. Roy, `Separability inequalities for N-Qudit systems
and their quantum violations', in preparation.
\end{enumerate} 

\end{document}